\documentclass[
 aip,
 amsmath,amssymb,
 reprint,%
]{revtex4-1}

\usepackage{graphicx}
\usepackage{dcolumn}
\usepackage{bm}
\usepackage{hyperref}

\usepackage[utf8]{inputenc}
\usepackage[T1]{fontenc}
\usepackage{mathptmx}
\usepackage{etoolbox}
\usepackage{xcolor}

\newcommand{\Google}{\affiliation{Google Quantum AI, Goleta CA 93117, USA}}

\begin{document}
\begin{title}


\title{Readout of a quantum processor with high dynamic range Josephson parametric amplifiers}

\author{Theodore White}
\thanks{Authors to whom correspondence should be addressed: tcwhite@google.com and ofernaaman@google.com}
 \Google
\author{Alex Opremcak}\Google
\author{George Sterling}\Google
\author{Alexander Korotkov}\Google
\author{Daniel Sank}\Google

\author{Rajeev Acharya}\Google
\author{Markus Ansmann}\Google
\author{Frank Arute}\Google
\author{Kunal Arya}\Google
\author{Joseph C.~Bardin}\Google
\affiliation{Department of Electrical and Computer Engineering, University of Massachusetts, Amherst MA, USA}
\author{Andreas Bengtsson}\Google
\author{Alexandre Bourassa}\Google
\author{Jenna Bovaird}\Google
\author{Leon Brill}\Google
\author{Bob B.~Buckley}\Google
\author{David A.~Buell}\Google
\author{Tim Burger}\Google
\author{Brian Burkett}\Google
\author{Nicholas Bushnell}\Google
\author{Zijun Chen}\Google
\author{Ben Chiaro}\Google
\author{Josh Cogan}\Google
\author{Roberto Collins}\Google
\author{Alexander L.~Crook}\Google
\author{Ben Curtin}\Google
\author{Sean Demura}\Google
\author{Andrew Dunsworth}\Google
\author{Catherine Erickson}\Google
\author{Reza Fatemi}\Google
\author{Leslie Flores~Burgos}\Google
\author{Ebrahim Forati}\Google
\author{Brooks Foxen}\Google
\author{William Giang}\Google
\author{Marissa Giustina}\Google
\author{Alejandro Grajales~Dau}\Google
\author{Michael C.~Hamilton}\Google
\affiliation{Department of Electrical and Computer Engineering, Auburn University, Auburn AL, USA}
\author{Sean D.~Harrington}\Google
\author{Jeremy Hilton}\Google
\author{Markus Hoffmann}\Google
\author{Sabrina Hong}\Google
\author{Trent Huang}\Google
\author{Ashley Huff}\Google
\author{Justin Iveland}\Google
\author{Evan Jeffrey}\Google
\author{M\'arika Kieferov\'a}\Google
\author{Seon Kim}\Google
\author{Paul V.~Klimov}\Google
\author{Fedor Kostritsa}\Google
\author{John Mark Kreikebaum}\Google
\author{David Landhuis}\Google
\author{Pavel Laptev}\Google
\author{Lily Laws}\Google
\author{Kenny Lee}\Google
\author{Brian J.~Lester}\Google
\author{Alexander Lill}\Google
\author{Wayne Liu}\Google
\author{Aditya Locharla}\Google
\author{Erik Lucero}\Google
\author{Trevor McCourt}\Google
\author{Matt McEwen}\Google
\affiliation{Department of Physics, University of California, Santa Barbara CA, USA}
\author{Xiao Mi}\Google
\author{Kevin C.~Miao}\Google
\author{Shirin Montazeri}\Google
\author{Alexis Morvan}\Google
\author{Matthew Neeley}\Google
\author{Charles Neill}\Google
\author{Ani Nersisyan}\Google
\author{Jiun How Ng}\Google
\author{Anthony Nguyen}\Google
\author{Murray Nguyen}\Google
\author{Rebecca Potter}\Google
\author{Chris Quintana}\Google
\author{Pedram Roushan}\Google
\author{Kannan Sankaragomathi}\Google
\author{Kevin J.~Satzinger}\Google
\author{Christopher Schuster}\Google
\author{Michael J.~Shearn}\Google
\author{Aaron Shorter}\Google
\author{Vladimir Shvarts}\Google
\author{Jindra Skruzny}\Google
\author{W.~Clarke Smith}\Google
\author{Marco Szalay}\Google
\author{Alfredo Torres}\Google
\author{Bryan Woo}\Google
\author{Z.~Jamie Yao}\Google
\author{Ping Yeh}\Google
\author{Juhwan Yoo}\Google
\author{Grayson Young}\Google
\author{Ningfeng Zhu}\Google
\author{Nicholas Zobrist}\Google

\author{Yu Chen}\Google
\author{Anthony Megrant}\Google
\author{Julian Kelly}\Google
\author{Ofer Naaman}
\thanks{Authors to whom correspondence should be addressed: tcwhite@google.com and ofernaaman@google.com}
\Google

\date{\today}

\begin{abstract}
We demonstrate a high dynamic range Josephson parametric amplifier (JPA) in which the active nonlinear element is implemented using an array of rf-SQUIDs. The device is matched to the 50 $\Omega$ environment with a Klopfenstein-taper impedance transformer and achieves a bandwidth of 250-300 MHz, with input saturation powers up to $-95$~dBm at 20 dB gain. A 54-qubit Sycamore processor was used to benchmark these devices, providing a calibration for readout power, an estimate of amplifier added noise, and a platform for comparison against standard impedance matched parametric amplifiers with a single dc-SQUID. We find that the high power rf-SQUID array design has no adverse effect on system noise, readout fidelity, or qubit dephasing, and we estimate an upper bound on amplifier added noise at 1.6 times the quantum limit. Lastly, amplifiers with this design show no degradation in readout fidelity due to gain compression, which can occur in multi-tone multiplexed readout with traditional JPAs.
\end{abstract}

\maketitle
\end{title}

Dispersive readout\cite{wallraff:dispersive} of superconducting qubits requires the use of near quantum limited superconducting amplifiers because of the severe limits placed by the quantum system on the allowed microwave probe power\cite{sank:readoutTransitions}. Resonant Josephson parametric amplifiers (JPA) have been popular in single qubit readout as they provide high gain, quantum-limited noise performance, tunable center frequency \cite{vijay2011observation, roch:JPC, JPADesign, aumentado2020superconducting}, and are simple to manufacture. However, larger quantum processors typically multiplex qubit measurement in the frequency domain, transmitting and receiving multiple probe tones using the same readout line\cite{google:quantsup}. This configuration requires a first stage amplifier with higher bandwidth and saturation power than the typical JPAs can provide.

\begin{figure}
\includegraphics{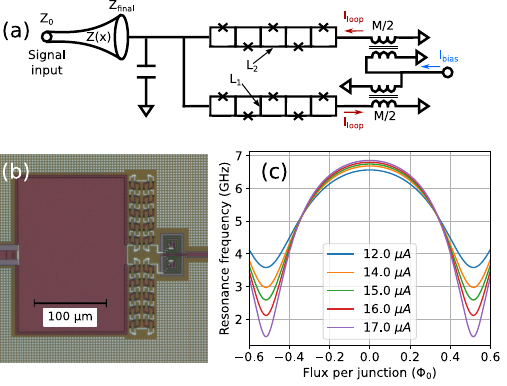}
\caption{\label{fig:circuit} (a) Circuit diagram of the amplifier.  The active element of the circuit consists of a tunable inductor made up of two rf-SQUID arrays in parallel (`snake').  Each rf-SQUID consists of a Josephson junction and geometric inductances $L_1$ and $L_2$, and each array contains 20 rf-SQUIDs.  A differential bias coil allows both DC and RF modulation of the inductance, while a shunt capacitor sets the frequency range.  A tapered impedance transformer lowers the loaded Q of the device, increasing its bandwidth. (b) Optical micrograph of the nonlinear resonator showing the shunt capacitor, snake inductor, and the bias transformer. (c) Calculated resonance frequency vs applied flux bias per junction (or per rf-SQUID), with $2N=40$ rf-SQUID stages, $C_s = 6.0$~pF, $L_1 = 2.6$~pH, $L_2 = 8.0$~pH, $L_b=30$~pH, and a range of junction $I_c$.}
\end{figure}

The instantaneous bandwidth of resonant JPAs can be increased using impedance matching techniques\cite{IMPA, vijay:broadband, duan2021broadband, naaman2022synthesis}, but their input saturation power remains low, $-115$~dBm to $-110$~dBm. High dynamic range JPAs using SQUID arrays have been demonstrated\cite{frattini2018optimizing, naaman:Snake}, but while improving the input saturation power up to $-95$~dBm, they still suffer from relatively narrow bandwidths, less than 200~MHz. Traveling wave parametric amplifiers (TWPA) can provide bandwidths in excess of 2 GHz and high  saturation power\cite{macklin:JTWPA, obrien:resTWPA, bockstiege:tiNTWPA2, eom:TiNparamp, esposito2021perspective}, but are difficult to fabricate and typically have lower quantum efficiency than JPAs due to higher dissipative and intermodulation losses\cite{peng2022floquet}. Practical systems considerations, such as qubit frequency placement plan\cite{barends:gates}, Purcell filter bandwidth\cite{jeffrey:readout}, and mixer IF bandwidth\cite{google:quantsup}, can additionally prevent the full utilization of the TWPAs multi-GHz bandwidth. 

Here, we demonstrate a resonant Josephson parametric amplifier that achieves the bandwidth performance of a matched JPA\cite{IMPA} and a hundred-fold increase in saturation power. The amplifier design is based on the impedance matched parametric amplifier (IMPA)\cite{IMPA}, which is in widespread use in our lab for frequency multiplexed readout. Unlike the IMPA, and indeed most JPA implementations, in which the amplifier's nonlinear inductance is provided by a single dc-SQUID (with critical currents of order of a few $\mu$A), the amplifiers presented in this Letter use arrays of high critical current rf-SQUIDs\cite{naaman:Snake}. The substitution of an rf-SQUID array for each of the junctions in the JPA SQUID increases the saturation power of the amplifier while keeping the total inductance of the device roughly the same.

Figure.~\ref{fig:circuit}(a) shows a simplified diagram of the amplifier. The nonlinear inductance is composed of two rf-SQUID arrays arranged in parallel to form a compound SQUID\cite{naaman:Snake}, with a total of $2N=40$ unit cells. Each rf-SQUID consists of a junction with a critical current $I_c$, and a linear inductance composed of two segments with inductance $L_1$ and one segment with inductance $L_2$. The $L_1$ segments are shared between neighboring rf-SQUIDs, so that the structure as a whole forms a serpentine inductive spine of alternating $L_1$ and $L_2$, bridged by Josephson junctions at each meander. We will refer to this structure as the `snake', and to the amplifier as a whole as `snake-IMPA` or `SNIMPA' for short. The parallel arrangement of the arrays enables us to flux-bias and pump the amplifier via a single superconducting split-coil spiral transformer as shown in the figure.  The transformer primary coils are connected in parallel paths between the bias feedline and ground. The secondary coils are counter-wound to primarily couple the bias to the circulating current mode ($I_\mathrm{loop}$ in the figure) of the compound SQUID loop formed by the two parallel portions of the snake (see Supplementary Material). The transformer mutual inductance is $M\approx50$~pH, and the self-inductance of the secondary coil is $\approx120$~pH. Finally, the structure is shunted by a capacitor $C_s$, and is connected to the $50\,\Omega$ signal port via a Klopfenstein taper as in Ref.~\onlinecite{IMPA}. An optical micrograph of the shunt capacitor, compound snake-SQUID, and bias transformer is shown in Fig.~\ref{fig:circuit}(b).

The resonance frequency of the nonlinear circuit created by the capacitively-shunted snake is $\omega_\mathrm{res} = 1 / \sqrt{(L_s+L_b)C_s}$, where $L_s$ is given by\cite{naaman:Snake, naaman2017josephson} (see Supplementary Material)
\begin{equation}
L_s = \frac{N}{2} \times \frac{L_J (L_1 + L_2) + L_1 L_2 \cos\delta_0}{L_J + (4L_1 + L_2)\cos\delta_0}, \label{snake2}
\end{equation}
$L_J=\hbar/2eI_c$, and $\delta_0$ is the equilibrium junction phase, which is dependent on the flux bias\cite{naaman:Snake, naaman2017josephson}. $L_b$ is a stray inductance that includes a contribution from the bias transformer self-inductance. The calculated resonance frequency for typical circuit parameters is shown in Fig.~\ref{fig:circuit}(c) as a function of applied flux bias per junction (or per rf-SQUID) in the array.

To avoid hysteresis in the modulation curve, the snake must be designed such that $L_J>4L_1+L_2$.  This means that, unlike a conventional dc-SQUID, the inductance of a snake-SQUID does not diverge at $\Phi_0 / 2$. Therefore $\omega_\mathrm{res}$ has a limited tunability range compared to conventional JPAs, as seen in Fig.~\ref{fig:circuit}(c). We therefore choose the shunt capacitance $C_s$ such that the frequency tunability range overlaps the desired operating frequency of the amplifier, 4.5-5.0 GHz. The device is flux-pumped at frequency $\omega_P$, which is twice the center frequency of the amplifier.

Here, we report on amplifiers having two design variants. The first, Design 1, has a nominal junction $I_c=16\,\mu$A and a shunt capacitance of $C_s=6.5$~pF, and the second, Design 2, has a nominal $I_c=18\,\mu$A and $C_s=6.0$~pF. Both designs have nominal snake inductances of $L_1=2.6$~pH and $L_2=8.0$~pH. From measurement of test junctions, we estimate that the actual critical current of the snake junctions is $\approx20$\% lower than designed. The Klopfenstein taper that matches the low impedance SNIMPA resonator to the $50\,\Omega$ signal port is similar to that reported in Ref.~\onlinecite{IMPA}, with a cutoff frequency of 2.6~GHz and a 50-section taper from $51\,\Omega$ to $24\,\Omega$, resulting in a resonator loaded Q of about 4.5. The devices were built in a three-layer aluminum process with SiO$_x$ interlayer dielectrics and Al/AlO$_x$/Al trilayer Josephson junctions. The first metal layer was used as a solid ground plane under the snake structure, and the upper two layers form the circuit elements of the amplifier (see Supplementary Material). In operation, the amplifiers' dc flux bias amounts to approximately $320\;\mu$A carried in each of the bias transformer primary coils, well below the measured critical current of traces and vias in our process.

\begin{figure}[h]
\includegraphics{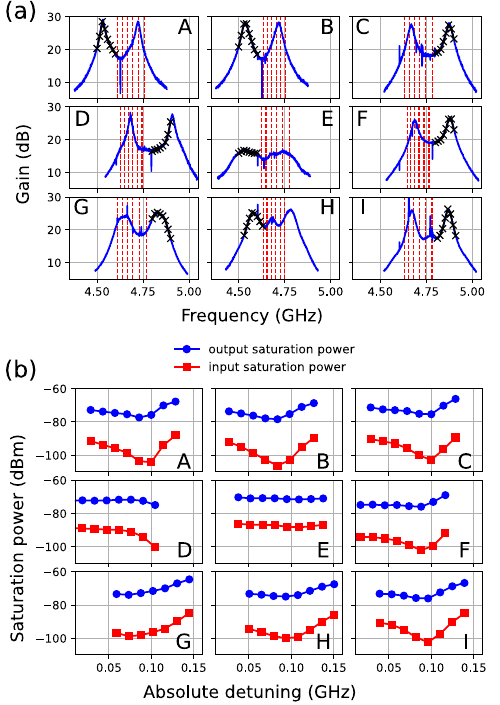}
\caption{\label{fig:performance} (a) Gain vs frequency (solid blue lines) of SNIMPA amplifiers, for each readout line as labeled. Lines A, C, E, and G use Design 1, and the rest use Design 2. Vertical lines (dashed red) indicate the frequencies of the readout resonators. Black crosses denote the frequencies at which saturation power was measured. (b) Corresponding input and output saturation power (1-dB gain compression), calibrated at the processor reference plane, vs absolute detuning from the amplifier center frequency.}
\end{figure}

\begin{figure}
\includegraphics[width=3.2in]{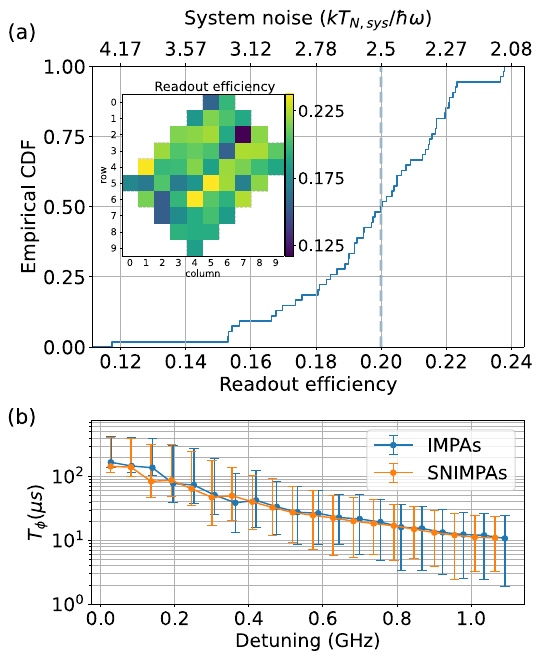}
\caption{\label{fig:comparison} (a) Cumulative distribution of readout efficiency over 54 qubits, showing a mean (dash-dot line) and median (dashed line) around 20 percent. The top x-axis represents the equivalent total system noise, expressed in terms of photon number units. Inset: heat-map of readout efficiency by qubit across the processor. (b) Dephasing times, $T_\phi$, measured using CPMG, vs qubit detuning away from their flux-insensitive, maximum frequency points. Data shows the median over all qubits in the processor, error bars represent the $17-83$ percentile range. Data measured with SNIMPA (orange) is compared with those measured with our standard IMPA (blue), showing similar performance of the two amplifier types.}
\end{figure}

We have characterized the performance of the SNIMPA devices with a 54-qubit Sycamore processor\cite{google:quantsup}. The processor consists of nine independent readout lines, each with six frequency-multiplexed readout resonators occupying a 4.6-4.8~GHz band. Each readout line has an on-chip Purcell filter\cite{jeffrey:readout}, and is connected to a SNIMPA amplifier through four circulators. The readout resonators and Purcell filters were designed with a target resonator ringdown time of $1/\kappa = 25$~ns. The readout lines are labeled A through I; lines A, C, E, and G had a SNIMPA with Design 1, and the rest had Design 2.  All SNIMPA were packaged with a magnetic shield. In a separate cooldown, all readout lines were outfitted with standard dc-SQUID based IMPA, whose performance we use as a baseline for comparison.

Figure.~\ref{fig:performance}(a) shows the SNIMPA gain vs signal frequency (blue lines) on all readout lines as labeled.  Each curve was measured at low power with a vector network analyzer,  after manual tuneup of the amplifier's flux bias, pump frequency, and pump power. The frequencies of the readout resonators associated with each line are indicated by the vertical dashed lines (red). All amplifiers achieve gains greater than 15 dB over the entire readout band, and most resonators can be read out with a gain exceeding 20 dB. The gain profile is not Lorentzian; this is because the complex impedance seen by the SNIMPA nonlinear resonator varies over the amplifier band\cite{IMPA}. The multi-peak response is commonly seen in wider band parametric amplifiers\cite{IMPA, grebel2021flux, ranzani2022wideband}, but impedance matching network synthesis techniques could be used to achieve better control of gain flatness and ripple\cite{naaman2022synthesis}.

For the purpose of the present experiments we were focused on isolated qubit readout performance. In some cases, we allowed the center frequency of the amplifier to reside within the resonator band (\textit{e.g.}~line H). In simultaneous multi-qubit readout with degenerate parametric amplifiers, in which the signal (at frequency $\omega_s$) and idler (at $\omega_i=\omega_P-\omega_s$) share the same physical circuit, this can cause interference between a signal from one of the resonators and an idler generated by the readout of another. Therefore, in simultaneous multi-qubit readout, the pump should be tuned such that all readout resonator frequencies reside either below or above $\omega_P/2$, meaning that only half the bandwidth is usefully available in practice. 

\begin{figure*}[ht]
\includegraphics{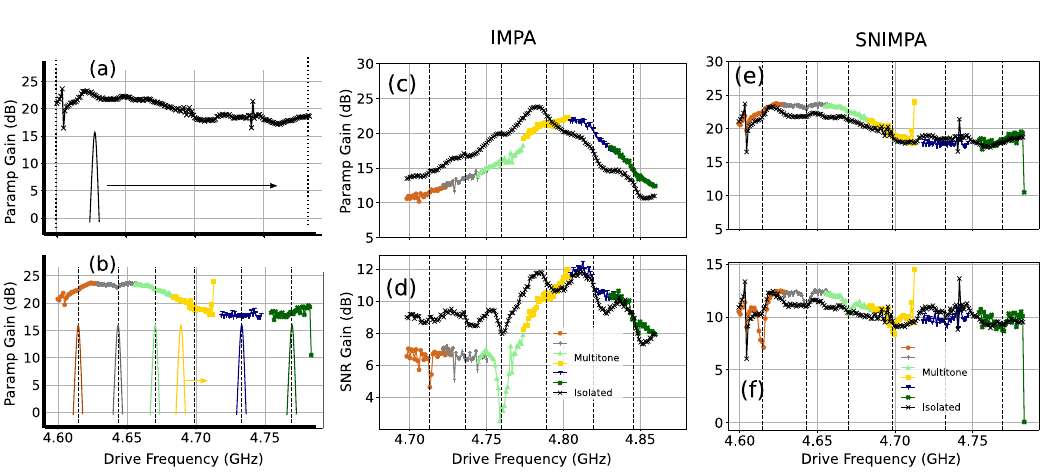}
\caption{\label{fig:multitone} Multi-tone readout experiments. (a)-(b) description of the experiment. (a) The amplifier gain is measured with a single tone at low power levels, typically -130 dBm, and the single tone is swept across the full readout band, and (b) gain is measured for each one of six readout tones as it is swept in turn across its corresponding frequency band (colored sections) while the other five tones are held at fixed frequencies as indicated by the respective vertical dashed lines. In this experiment each of the six tones are roughly -120 dBm, i.e.~10~dB higher power per tone than the baseline.  (c) Measured power gain, and (d) SNR gain in the multi-tone experiment (color, `multitone') compared to baseline (black, `isolated') for a traditional IMPA amplifier. The IMPA shows a shift in the gain profile and a degradation of the SNR gain due to saturation. (e)-(f) Results of the same experiment using the SNIMPA amplifier on readout line G, which shows little change in power gain and SNR gain when driven simultaneously with six high power tones, as compared to the single tone low power baseline.}
\end{figure*}

Figure \ref{fig:performance}(b) shows the input (squares) and output (circles) saturation power (1-dB gain compression) of each of the amplifiers, as a function of absolute signal detuning from the amplifier center frequency (half the pump frequency, $\omega_P/4\pi$).  The crosses in Fig.~\ref{fig:performance}(a) denote the frequency and gain at which each point was measured. For this experiment we calibrate the signal power at the processor reference plane using ac Stark shift of the processor qubits. We first measure the dispersive shift $\chi=(\omega_{r,|0\rangle}-\omega_{r,|1\rangle})/2$ (where $\omega_{r,|0\rangle}$ and $\omega_{r,|1\rangle}$ are the dressed resonance frequencies), and the resonator decay rate $\kappa$ spectroscopically for each qubit in our processor. From the ac Stark shift, which is given (in the linear regime) by $\delta\omega_{01} = -2 \chi \bar{n}$, we extract $\bar{n}$, the average resonator photon occupation for a given combination of resonator drive power and frequency\cite{Schuster2007}. From the measured values of $\kappa$ and $\bar{n}$ we calculate the microwave power impinging on the resonator\cite{vijay2012stabilizing}. Repeating this measurement at varying powers from the room-temperature signal generator allows us to calibrate the power delivered to the chip as a function of signal generator power. Figure~\ref{fig:performance}(b) shows that the SNIMPA consistently achieve output saturation powers in the $-80$~dBm to $-70$~dBm range, up to 20 dB higher than our standard dc-SQUID based IMPA. While output saturation power, being ideally gain-independent, is the more fundamental quantity, the figure also reports the input saturation power to allow a more direct comparison with the bulk of the existing literature. 

We do not measure the amplifier added noise directly. Instead, we focus on the overall readout efficiency\cite{QuantEff}, $\eta$, which is a more relevant metric from a systems performance perspective (see Supplementary Material). Readout efficiency encapsulates all microwave losses, $\alpha$, between the readout resonators and the SNIMPA, as well as the noise temperature of the SNIMPA itself, $T_p$, and the effective noise temperature, $T_h$, of the cryogenic HEMT amplifier and the rest of the measurement chain following the SNIMPA.  Since our amplifiers are operated in a phase preserving mode, the maximum possible efficiency is $\eta=0.5$. 

Fig.~\ref{fig:comparison}(a) shows the measured readout efficiency for all qubits in our Sycamore processor.  The empirical cumulative distribution function of these data is shown in the main plot, with a median of $\eta=0.2$ corresponding to a system noise temperature of $T_{N,\mathrm{sys}}=560$~mK.  The inset shows how these efficiencies are distributed across the qubit grid.  

We measure the readout efficiency, $\eta$, and signal-to-noise ratio (SNR) gain, $G_\mathrm{SNR}$, as a function of SNIMPA power gain, $G_p$, for all qubits in our processor:
\begin{align}
    G_\mathrm{SNR} &= \frac{G_p\left(T_Q+T_h\right)}{G_pT_Q+\left(G_p-1\right)T_p+T_h} \label{eq:snr_gain}\\
    \eta &= \frac{\alpha G_pT_Q}{G_pT_Q+\left(G_p-1\right)T_p+T_h}, \label{eq:efficiency}
\end{align}
where $T_Q=\hbar\omega/2k$ is the quantum noise at the readout frequency, and fit the data with both equations simultaneously. Since these fits cannot separate out contributions from $\alpha$ and $T_p$, we have to fix one of these parameters. If the SNIMPA were quantum limited and we fix $T_p=T_Q$, then fits to our data suggest an average insertion loss of $\alpha=0.44\pm0.03$, or $-3.57$~dB, between the processor chip and the SNIMPA. 

Independent, cryogenically calibrated\cite{wang2021cryogenic} measurements of individual component losses (circulators and wiring) add up to a minimum of $-2.5$~dB of loss between the processor chip and the SNIMPA (see Supplementary Material). If we therefore fix $\alpha=0.56$ ($-2.5$~dB) in the fits, then we can extract a maximum SNIMPA noise temperature, $T_p=0.18\pm0.02$~K, roughly $60\%$ higher than the quantum limit. A cryogenically calibrated measurement of a representative integrated readout assembly shows an insertion loss that varies between $-3.6$~dB and $-2.7$~dB over the readout band (see Supplementary Material). These data put bounds on the average noise performance of the SNIMPA, as deployed, in the context of qubit readout of our processor. The data suggest that after accounting for component losses, the measured efficiencies are consistent with near quantum limited noise performance of the SNIMPA amplifiers.

A potential concern with the SNIMPA is that the snake inductor could increase back action on the qubit, compared to the standard dc-SQUID based amplifiers.  This could be due to coupling of noise photons through the large bias transformer, pump leakage to the signal line, or the generation of spurious signals or noise during amplification. Figure~\ref{fig:comparison}(b) shows qubit dephasing time, $T_\phi$, as measured using the CPMG (Carr-Purcell-Meiboom-Gill) method\cite{LL:CPMG}, vs qubit detuning away from their flux-insensitive point. The data, representing the median over all qubits in the processor, compare the performance measured with the SNIMPAs (orange) to that measured in a separate cooldown of the same processor with the standard IMPA amplifiers (blue). The data indicate that the SNIMPA have no adverse effect on qubit dephasing as compared to our standard dc-SQUID based amplifiers.

Finally, we performed multi-tone experiments to test the SNIMPA performance in an emulated simultaneous multi-qubit readout scenario. We chose to do so, instead of directly characterizing multi-qubit readout fidelity, to avoid confounding qubit-related physics at high readout powers\cite{sank:readoutTransitions} that can mask the underlying performance of the amplifiers. As a baseline, we first measure the amplifier signal power gain and SNR improvement (SNR gain) as a function of frequency with a single (`isolated'), weak readout tone (approximately $-130$~dBm), as shown in Fig.~\ref{fig:multitone}(a). We then repeat the measurement with five additional tones (emulating the six-qubit multiplexed readout in the Sycamore processor\cite{google:quantsup}), and with each tone having 10 dB higher power, as illustrated in Fig.~\ref{fig:multitone}(b). Here, each one of the readout tones is swept in turn across its corresponding frequency band (respective color) and we measure the signal and SNR gains while the others tones are on and are kept at the fixed frequencies indicated by the respective vertical dashed lines. Figure~\ref{fig:multitone}(c) and (d) show the gain and SNR improvement, respectively, for our standard dc-SQUID based IMPA, clearly showing a degradation in both quantities in the multi-tone experiment (color) compared with the low-power baseline (black). In contrast, no such degradation is observed with the SNIMPA, as shown in Fig.~\ref{fig:multitone}(e)-(f). In fact, we found it difficult to drive the SNIMPA to saturation with our standard readout electronics setup. These experiments demonstrate that the SNIMPA's high saturation power offers sufficient headroom in a multi-qubit readout situation to enable more efficient readout multiplexing with higher number of qubits read-out simultaneously per channel. The SNIMPA can also accommodate higher power per readout tone, which may be required for high-fidelity, fast readout with greater qubit-resonator detuning or weaker qubit-resonator coupling.

In summary, we have demonstrated impedance matched Josephson parametric amplifiers with saturation powers up to two orders of magnitude higher than their standard dc-SQUID based counterparts. By combining an rf-SQUID array nonlinear element with an impedance matching taper, these amplifiers achieve sufficient instantaneous bandwidth to support 6:1 frequency-multiplexed readout of our Sycamore processors. We have measured a median readout efficiency of 20\%, extracted an upper bound for the SNIMPA noise temperature at around 60\% over the quantum limit, and found no excess amplifier-related dephasing compared to our standard IMPA-based setup. 

\section{Supplementary Material}
See supplementary material for details on the derivation of the snake inductance, transformer design, efficiency measurement, and loss budget estimates.
%

\clearpage

\end{document}


\title{Supplementary information: Readout of a quantum processor with high dynamic range Josephson parametric amplifiers}
\author{Theodore White}
\thanks{Authors to whom correspondence should be addressed: tcwhite@google.com and ofernaaman@google.com}
 \Google
\author{Alex Opremcak}\Google
\author{George Sterling}\Google
\author{Alexander Korotkov}\Google
\author{Daniel Sank}\Google

\author{Rajeev Acharya}\Google
\author{Markus Ansmann}\Google
\author{Frank Arute}\Google
\author{Kunal Arya}\Google
\author{Joseph C.~Bardin}\Google
\affiliation{Department of Electrical and Computer Engineering, University of Massachusetts, Amherst MA, USA}
\author{Andreas Bengtsson}\Google
\author{Alexandre Bourassa}\Google
\author{Jenna Bovaird}\Google
\author{Leon Brill}\Google
\author{Bob B.~Buckley}\Google
\author{David A.~Buell}\Google
\author{Tim Burger}\Google
\author{Brian Burkett}\Google
\author{Nicholas Bushnell}\Google
\author{Zijun Chen}\Google
\author{Ben Chiaro}\Google
\author{Josh Cogan}\Google
\author{Roberto Collins}\Google
\author{Alexander L.~Crook}\Google
\author{Ben Curtin}\Google
\author{Sean Demura}\Google
\author{Andrew Dunsworth}\Google
\author{Catherine Erickson}\Google
\author{Reza Fatemi}\Google
\author{Leslie Flores~Burgos}\Google
\author{Ebrahim Forati}\Google
\author{Brooks Foxen}\Google
\author{William Giang}\Google
\author{Marissa Giustina}\Google
\author{Alejandro Grajales~Dau}\Google
\author{Michael C.~Hamilton}\Google
\affiliation{Department of Electrical and Computer Engineering, Auburn University, Auburn AL, USA}
\author{Sean D.~Harrington}\Google
\author{Jeremy Hilton}\Google
\author{Markus Hoffmann}\Google
\author{Sabrina Hong}\Google
\author{Trent Huang}\Google
\author{Ashley Huff}\Google
\author{Justin Iveland}\Google
\author{Evan Jeffrey}\Google
\author{M\'arika Kieferov\'a}\Google
\author{Seon Kim}\Google
\author{Paul V.~Klimov}\Google
\author{Fedor Kostritsa}\Google
\author{John Mark Kreikebaum}\Google
\author{David Landhuis}\Google
\author{Pavel Laptev}\Google
\author{Lily Laws}\Google
\author{Kenny Lee}\Google
\author{Brian J.~Lester}\Google
\author{Alexander Lill}\Google
\author{Wayne Liu}\Google
\author{Aditya Locharla}\Google
\author{Erik Lucero}\Google
\author{Trevor McCourt}\Google
\author{Matt McEwen}\Google
\affiliation{Department of Physics, University of California, Santa Barbara CA, USA}
\author{Xiao Mi}\Google
\author{Kevin C.~Miao}\Google
\author{Shirin Montazeri}\Google
\author{Alexis Morvan}\Google
\author{Matthew Neeley}\Google
\author{Charles Neill}\Google
\author{Ani Nersisyan}\Google
\author{Jiun How Ng}\Google
\author{Anthony Nguyen}\Google
\author{Murray Nguyen}\Google
\author{Rebecca Potter}\Google
\author{Chris Quintana}\Google
\author{Pedram Roushan}\Google
\author{Kannan Sankaragomathi}\Google
\author{Kevin J.~Satzinger}\Google
\author{Christopher Schuster}\Google
\author{Michael J.~Shearn}\Google
\author{Aaron Shorter}\Google
\author{Vladimir Shvarts}\Google
\author{Jindra Skruzny}\Google
\author{W.~Clarke Smith}\Google
\author{Marco Szalay}\Google
\author{Alfredo Torres}\Google
\author{Bryan Woo}\Google
\author{Z.~Jamie Yao}\Google
\author{Ping Yeh}\Google
\author{Juhwan Yoo}\Google
\author{Grayson Young}\Google
\author{Ningfeng Zhu}\Google
\author{Nicholas Zobrist}\Google

\author{Yu Chen}\Google
\author{Anthony Megrant}\Google
\author{Julian Kelly}\Google
\author{Ofer Naaman}
\thanks{Authors to whom correspondence should be addressed: tcwhite@google.com and ofernaaman@google.com}
\Google

\date{\today}
\maketitle

\section{Analysis of the Snake Inductance}\label{apx:snake}

\begin{figure}
\includegraphics[width=\columnwidth]{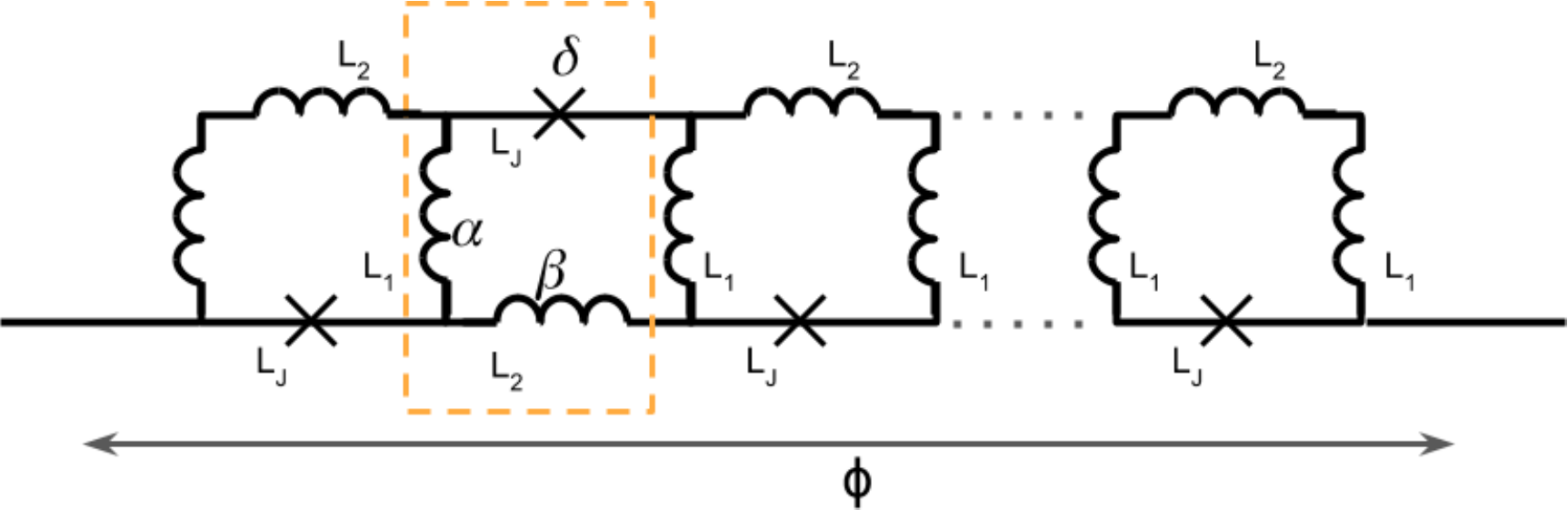}
\caption{\label{fig:snake_schem}Schematic of the snake array, with the phases $\alpha$, $\beta$, and $\delta$ indicated. The unit cell is highlighted with a dashed box.}
\end{figure}

Here, we outline the derivation of the snake SQUID array potential, current-phase relation, and inductance. A schematic of the snake array is shown in Fig.~\ref{fig:snake_schem}. We will assume that the array is sufficiently long, so that we can neglect edge effects, and treat the array as translationally invariant\cite{bell2012quantum}. 

\begin{figure}
\includegraphics[width=1.0in]{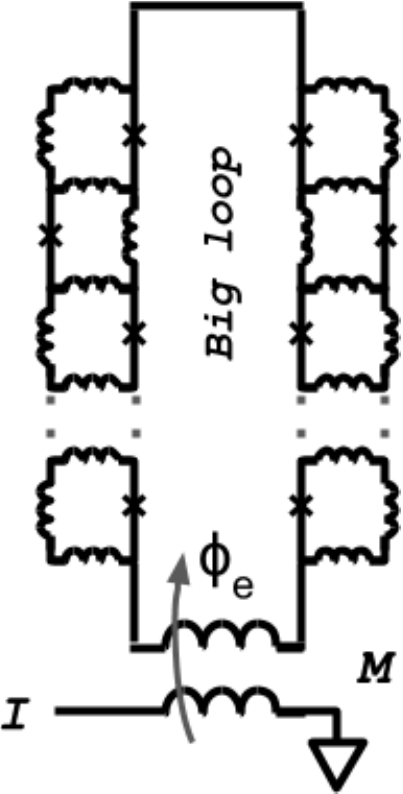}
\caption{\label{fig:snake_squid}Schematic of the snake SQUID, flux-biased through a transformer with self-inductance of $L_b$ and mutual $M$ such that $\phi_e=(2\pi/\Phi_0)\times MI$}
\end{figure}

We will refer to the inductive loops containing a junction and inductors $L_1$ and $L_2$, essentially an rf-SQUID, as the ``small loop" (see Fig.~\ref{fig:snake_schem}). In the small loop, the phase drop across $L_1$ is $\alpha$ and the phase drop across $L_2$ is $\beta$. The junction phase is $\delta$ as usual. The sign convention for the phases is such that $\beta$ and $\delta$ grow from left to right, whereas the direction of $\alpha$ alternates downwards and upwards as the inductive spine of the snake meanders. We will refer to the loop that encloses the full snake array and any inductance $L_b$ external to it (for example, the self-inductance of a bias transformer) as the ``big loop", as shown in Fig.~\ref{fig:snake_squid}. The snake we will consider has $2N$ unit cells, so when implementing a ``dc-SQUID-like" structure (Fig.~\ref{fig:snake_squid}) we will have two $N$-unit-cell arrays in parallel.

From flux quantization of any of the small loops of the snake we have the relation between the phases:
\begin{equation}\label{small_loop}
2\alpha+\beta=\delta,
\end{equation}
and with a total phase drop of $\phi$ across the full array with 2$N$ sections array we have:
\begin{equation}\label{total_phase}
2N\left(\delta-\alpha\right)=\phi.
\end{equation}

We can write the total potential energy of the array and use Eqs.~(\ref{small_loop}) and (\ref{total_phase}) to eliminate $\alpha$ and $\beta$:
\begin{align}\label{potential}
U_s &= 2N\left(\frac{\Phi_0}{2\pi}\right)^2\times\left(\frac{\alpha^2}{2L_1}+\frac{\beta^2}{2L_2}-\frac{1}{L_J}\cos\delta\right)\\
&=2N\left(\frac{\Phi_0}{2\pi}\right)^2\times\left[\frac{\left(\delta-\frac{\phi}{2N}\right)^2}{2L_1}+\frac{\left(\frac{\phi}{N}-\delta\right)^2}{2L_2}-\frac{1}{L_J}\cos\delta\right]
\end{align} 

So far this is what has been done in Ref.~\onlinecite{naaman2017josephson}, and follows the reasoning in Ref.~\onlinecite{bell2012quantum}. Here we want to also include the effects of an inductor $L_b$ closing the loop, through which we flux-bias the big loop with flux $\left(\frac{\Phi_0}{2\pi}\right)\phi_e$ as in Fig.~\ref{fig:snake_squid}. When doing so we need to account also for the potential energy stored in the inductor $L_b$, and we use flux quantization around the big loop to get:
\begin{equation}
U_b = \left(\frac{\Phi_0}{2\pi}\right)^2\frac{\left(\phi_e-\phi\right)^2}{2L_b},
\end{equation}
and $U_{tot}=U_s+U_b$.

From here we want to find the equilibrium phase $\delta_0$ that minimizes the potential. We set $\frac{dU}{d\delta}=0$ and $\frac{dU}{d\phi}=0$, and eliminate $\phi$ from the two resulting equations to find $\delta_0$ in terms of the external flux $\phi_e$.
The derivative with respect to $\delta_0$ gives the expression:
\begin{equation}\label{deriv_delta}
\delta_0\left(\frac{1}{L_1}+\frac{1}{L_2}\right)+\frac{1}{L_J}\sin\delta_0=\frac{\phi}{2N}\left(\frac{1}{L_1}+\frac{2}{L_2}\right),
\end{equation}

and the derivative with respect to $\phi$ gives:
\begin{equation}\label{deriv_phi}
\frac{\phi}{2N}=\frac{\frac{\phi_e}{L_b}+\delta\left(\frac{1}{L1}+\frac{2}{L_2}\right)}{\frac{1}{L_1}+\frac{4}{L_2}+\frac{2N}{L_b}}
\end{equation}

Using Eq.~(\ref{deriv_delta}) and Eq.~(\ref{deriv_phi}), and some algebra we finally get an implicit equation for $\delta_0(\phi_e)$:
\begin{widetext}
\begin{equation}\label{equilib}
\delta_0\left(L_1+L_2-L_b\frac{\left(2L_1+L_2\right)^2}{L_b\left(4L_1+L_2\right)+2NL_1L_2}\right)+\frac{L_1L_2}{L_J}\sin\delta_0=\phi_e\frac{L_1L_2\left(2L_1+L_2\right)}{L_b\left(4L_1+L_2\right)+2NL_1L_2}
\end{equation}
\end{widetext}
Note that when $L_b=0$, this equation reduces to Eq.~(\ref{deriv_delta}) as it should, where $\phi_e=\phi$. Eq.~(\ref{equilib}) can be solved numerically to find the equilibrium phase $\delta_0$ as a function of applied flux $\phi_e$.

Another way to get the same result is to consider the current-phase relation $I(\phi)$ of the snake itself, which can be found from the derivative of the potential with respect to the phase drop $\phi$:

\begin{align}\label{current-phase}
I&=\left. \left(\frac{2\pi}{\Phi_0}\right)\frac{dU}{d\phi}\right\rvert_{\delta_0}\\
&=\left(\frac{\Phi_0}{2\pi}\right)\times\left[-\delta_0\left(\frac{1}{L_1}+\frac{2}{L_2}\right)+\frac{\phi}{2N}\left(\frac{1}{L_1}+\frac{4}{L_2}\right)\right]
\end{align}

The external flux that needs to be applied to the ``big loop" to affect a phase drop $\phi$ across the array is $\frac{\Phi_0}{2\pi}\phi_e$, where
\begin{equation}
\phi_e=\phi+\frac{2\pi}{\Phi_0}L_bI(\phi),
\end{equation}
and using Eq.~(\ref{current-phase}):
\begin{equation}
\phi_e=\phi-L_b\left[\delta_0\left(\frac{1}{L_1}+\frac{2}{L_2}\right)-\frac{\phi}{2N}\left(\frac{1}{L_1}+\frac{4}{L_2}\right)\right],
\end{equation}
Which is equivalent to Eq.~(\ref{deriv_phi}).

The inductance of the snake SQUID, with the array tapped in the middle to form two $N$-section segments in parallel, and closed in the ``big loop" through $L_b$ can be found by taking the second derivative of the potential of Eq.~(\ref{potential}) with respect to the phase $\phi$, and since $\delta_0$ depends on $\phi$ as well, we have to be careful to capture the full derivative:

\begin{equation}
L_s=\frac{N}{2}\left(\frac{\Phi_0}{2\pi}\right)^2\times\left(\frac{\partial^2U}{\partial\phi^2}+\frac{\partial^2U}{\partial\phi\partial\delta_0}\frac{\partial\delta_0}{\partial\phi}\right)^{-1}.
\end{equation}
Carrying out the derivatives we get:
\begin{equation}
L_s=\frac{N}{2}\frac{\left(L_1+L_2\right)L_J+L_1L_2\cos\delta_0}{L_J+\left(4L_1+L_2\right)\cos\delta_0}.
\end{equation}

This is the contribution of the snake itself, with $\delta_0$ given by Eq.~(\ref{equilib}), and we can add $L_b/2$ to each of the SQUID parallel branches.

\begin{figure}[h]
\includegraphics[width=\columnwidth]{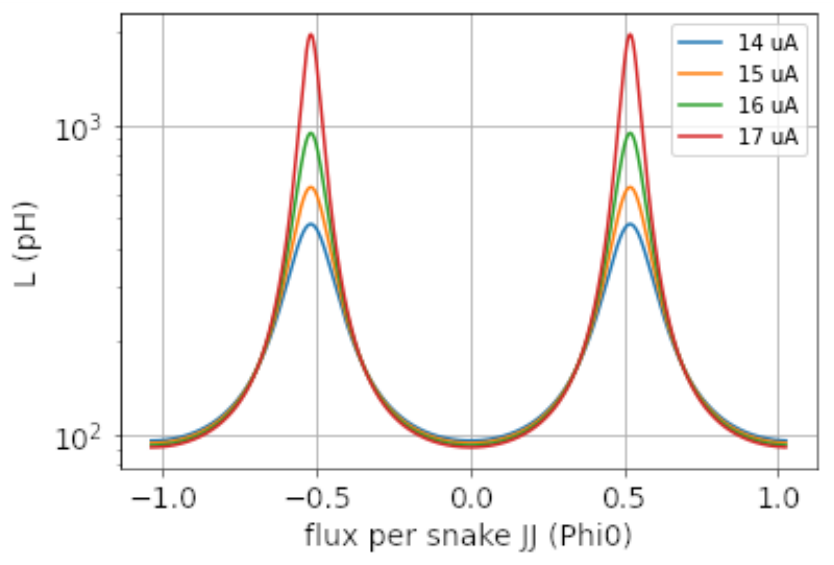}
\caption{\label{inductance}Inductance of the Snake SQUID, on a logarithmic scale, as a function of applied flux (per junction) for several junction $I_c$}
\end{figure}

Figure \ref{inductance} shows the calculated inductance vs flux curves for a device with $N=20$ junctions per arm, small-loop inductances $L_1=2.6$ pH and $L_2=8.0$ pH, $L_b=124$ pH, and for several values of $I_c$. We see that the inductance here is always positive (unlike that of the simple rf-SQUID), and that the depth of inductance modulation depends on the junction critical current, with lower $I_c$ resulting in shallower modulation.

Note that the applied flux in Fig.~\ref{inductance} is per junction, or in other words, a snake with $N$ junctions will require $N$ times more flux bias than a single rf-SQUID to reach the operating point. This is why our amplifier requires a large bias transformer, and in general higher pump currents than the traditional dc-SQUID based JPA.

\begin{figure}[th]
\includegraphics[width=3.1in]{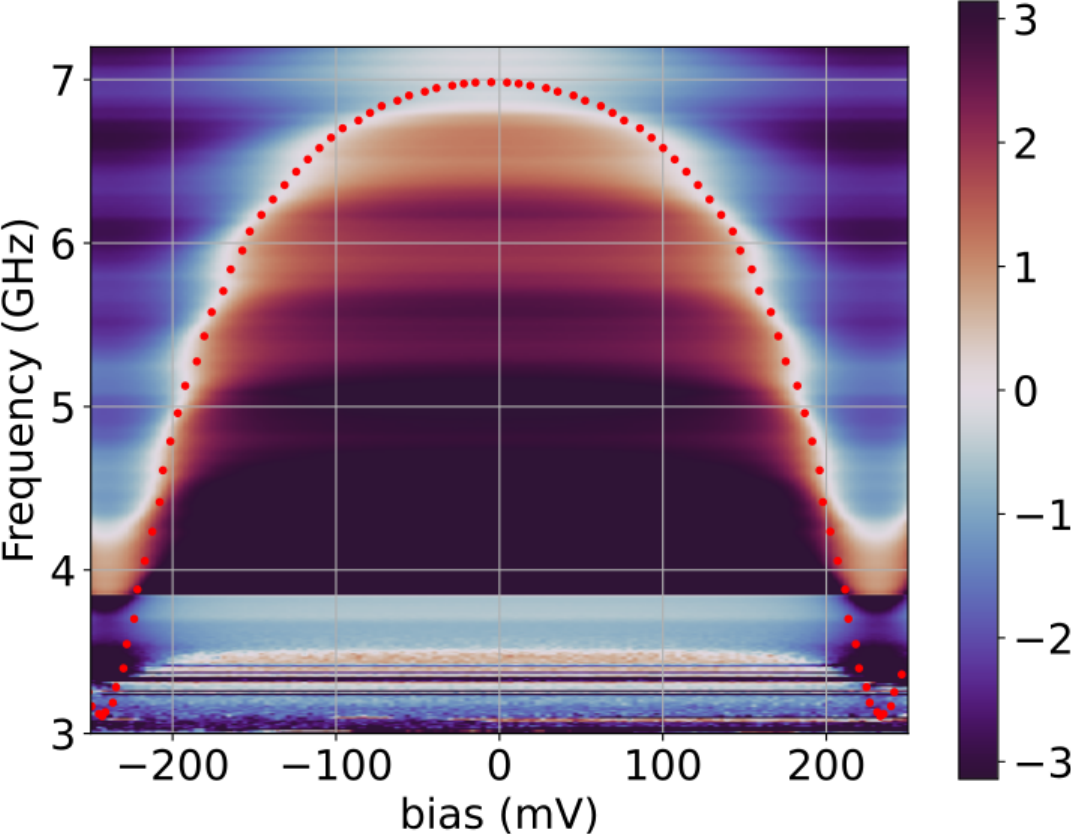}
\caption{\label{fig:s11_phase} Measured $S_{11}$ phase (rad) vs bias and frequency, for a SNIMPA amplifier with the pump turned off. The bias (shown as voltage sourced at the room-temperature electronics) was provided to the chip via a 20 dB attenuator at 3 K. Dotted red line indicates the calculated resonance frequency with $I_c=14\;\mu$A, $C_s=5.5$~pF, and with $L_b=120$~pH. Data at frequencies below 4 GHz are affected by the band edge of our circulators.}
\end{figure}

Figure~\ref{fig:s11_phase} shows the reflected phase vs bias from a SNIMPA amplifier similar to those discussed in the main text but with a smaller shunt capacitor, measured with the pump off. The dotted red curve in the figure indicates the calculated resonance frequency of the capacitively shunted snake inductor, using $I_c=14\;\mu$A, $C_s=5.5$~pF, and with a stray inductance of 30~pH ($L_b=120$~pH). The figure indicates that the inductance of the snake SQUID modulates as expected with flux bias, with negligible flux offset and a symmetric response about zero flux.

We found that magnetic shielding of the amplifiers is important to their operation. Unshielded devices show low gains, typically only 5-10 dB, with significant flux offset  and notable asymmetry in their modulation curves.  Note that the biasing scheme shown in Fig.~\ref{fig:snake_squid} results in neighboring rf-SQUIDs being biased in antiferromagentic fashion. Flux trapped in the array or its immediate vicinity, as well as any global magnetic field background, will therefor cause a non-uniform biasing of the device. 

\section{Bias transformer design}

The devices were built in a three-layer aluminum process with SiO$_x$ interlayer dielectrics and Al/AlO$_x$/Al trilayer Josephson junctions. One of the metal layers was used as a solid ground plane under the snake structure.

\begin{figure}[h]
\includegraphics[width=\columnwidth]{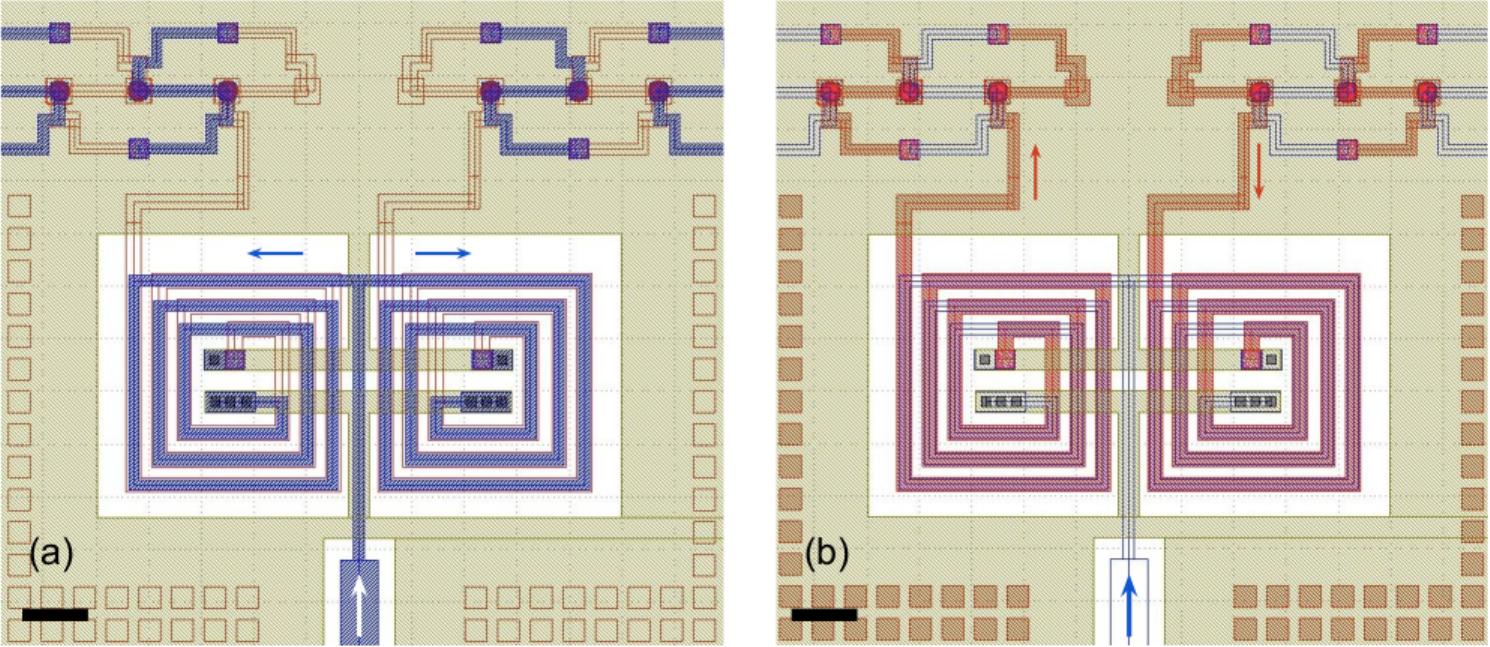}
\caption{\label{fig:xformer} Layout of the bias transformer and a portion of the snake array. Josephson junctions are shown as red circles along the center line of the arrays. (a) Primary bias coil highlighted in blue, (b) secondary coil highlighted in red. Scale bar is $6\;\mu$m.}
\end{figure}

Figure~\ref{fig:xformer} shows the layout of the bias transformer and three of the rf-SQUID stages in the two parallel snake arrays. The two panels show the same layout, but panel (a) highlights the primary split-coil of the transformer and the upper routing layer of the snake circuit (blue), and panel (b) highlights the secondary counter-wound coil and the lower routing layer (red) of the circuit. The bias and pump are supplied from the bottom of the figure, as indicated by the arrow. The groundplane is shown in tan color, and the snake's Josephson junctions are shown as red circles. 

In panel (a) the direction of current flow in the primary is indicated by the blue arrows in response to the input (white arrow), and the induced current in the secondary is shown in panel (b) by the red arrows. Both coils return to ground through the bars in the middle of the respective figures.

We have simulated the transformer using FastHenry\cite{fasthenry}. We estimate the mutual coupling between the bias line and the snake's circulating current mode to be approximately 51~pH, whereas the parasitic coupling from the bias line to the snake's signal mode is estimated to be approximately 1.9 ~pH.

\section{Readout efficiency measurement}\label{apx:efficiency}
Readout efficiency quantifies the amount of information about the qubit state that can be learned per readout photon\cite{QuantEff, vijay2012stabilizing}. In practice, this translates to the rate at which the readout signal-to-noise ratio (SNR) improves as a function of measurement time, $t$. Ideally, this quantity, which we call the SNR flux, is given by:
\begin{equation}\label{eq:ideal_snr}
    \Phi_\mathrm{SNR}^\mathrm{ideal}\equiv\left.\frac{d\text{SNR}}{dt}\right|_\mathrm{ideal}=8\bar{n}\kappa \frac{(2\chi/\kappa)^2}{1+\left(2\chi/\kappa\right)^2},
\end{equation}
where we have included half a photon of noise associated with the input measurement tone, and $\bar{n}$ is evaluated at the readout frequency, $\omega_\text{ro} = (\omega_{r,|0\rangle}+\omega_{r,|1\rangle})/2$, so that both dressed resonator states have the same value of $\bar{n}$.

In reality, not all photons are detected (due to microwave losses) and our phase-preserving amplification chain necessarily adds at least half a photon of noise so in an experiment the SNR flux is smaller than the ideal value by at least a factor of 2. We determine the readout efficiency from the ratio between the experimental and ideal SNR flux Eq.~(\ref{eq:ideal_snr}), $\eta=\Phi_\mathrm{SNR}/\Phi_\mathrm{SNR}^\mathrm{ideal}\le0.5$.

In the experiment, we record the readout SNR vs measurement time, which is a linear function for sufficiently low readout power, with a slope of $\Phi_\mathrm{SNR}$. Finally, using the measured values of $\chi$, $\kappa$, and $\bar{n}$, we calculate the experimental readout efficiency from the slope of a line fit to the data:
\begin{equation}
    \eta=\frac{\Phi_\mathrm{SNR}}{\Phi_\mathrm{SNR}^\mathrm{ideal}}= \frac{d\text{SNR}}{dt}\times\frac{1}{8\bar{n}\kappa}\left[\frac{1+\left(2\chi/\kappa\right)^2}{\left(2\chi/\kappa\right)^2}\right].
\end{equation}

\section{Loss measurement}\label{aps:loss}

We have used a cryogenic microwave calibration platform to characterize losses in our readout assembly. The measurement system is configured according to Ref.~\onlinecite{wang2021cryogenic}, and expanded to allow 2-port SOLT calibration\cite{ranzani2013two}.

We first characterize all components in our readout assembly separately, including cables, circulators, and circuit board traces. We can add up the insertion loss measured for these components to estimate a lower limit on the total loss in our system between the processor and the SNIMPA, $-2.5$~dB for the experiments described here.

\begin{figure}[h]
\includegraphics[width=\columnwidth]{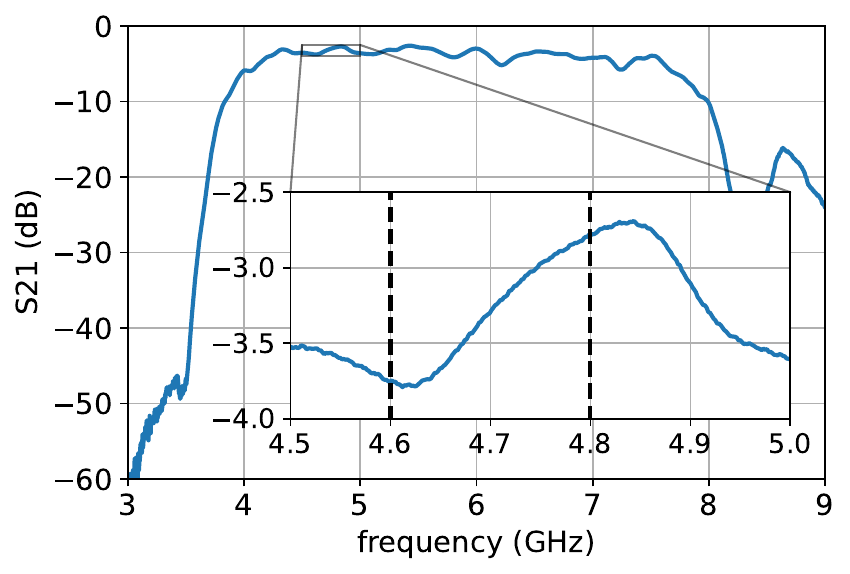}
\caption{\label{fig:loss} Cryogenically calibrated $S_{21}$ measurement of a representative integrated readout assembly, including all components from the processor chip to the amplifier chip. Inset: zoomed in view of the insertion loss in the readout band, indicated by the dashed vertical lines.}
\end{figure}

When components are cascaded and built into a complete readout assembly, reflections from circulators and connectors can contribute to the overall insertion loss. In addition, electrical delays associated with the electrical lengths of connectors and circuit board traces can add significant frequency dependence to the overall loss. Figure~\ref{fig:loss} shows a cryogenically calibrated $S_{21}$ measurement of a representative integrated readout assembly, similar to that used in the experiment, deembedded from the processor chip reference plane to the signal port of the SNIMPA. The circulators' 4-8 GHz band-pass response can be seen, with an average insertion loss around $-3$~dB. The inset shows a zoom-in on the frequency range corresponding to our readout band (highlighted with dashed lines in the inset), and we can see that indeed the insertion loss varies with frequency. 

The details of the insertion loss will vary from one readout line to the next, therefore in the main text we only use the measured minimum loss number, and obtain an upper bound on the SNIMPA added noise. However, as Fig.~\ref{fig:loss} suggests, the total insertion loss is likely higher, and the SNIMPA noise performance is likely consistent with near quantum limited operation.

%

\clearpage